\documentclass[12pt]{article}

\newcommand\independent{\protect\mathpalette{\protect\independenT}{\perp}}
\def\independenT#1#2{\mathrel{\rlap{$#1#2$}\mkern2mu{#1#2}}}

\usepackage{amsmath,amsfonts, amssymb, xcolor}
\usepackage{color}
\usepackage{graphicx}
\usepackage[margin=1.0in,paperheight=11in,paperwidth=8.5in]{geometry}
\usepackage{tikz}
\usepackage{pgfplots}
\usepackage{mathtools}
\usepackage{subcaption}
\usepackage{setspace}
\usepackage[superscript]{cite}
\usepackage{listings}
\usepackage{ulem}

\usetikzlibrary{arrows,shapes.arrows,shapes.geometric,shapes.multipart, decorations.pathmorphing,positioning, shapes.swigs,}

\makeatletter
\newcommand*{\centernot}{%
  \mathpalette\@centernot
}
\def\@centernot#1#2{%
  \mathrel{%
    \rlap{%
      \settowidth\dimen@{$\m@th#1{#2}$}%
      \kern.5\dimen@
      \settowidth\dimen@{$\m@th#1=$}%
      \kern-.5\dimen@
      $\m@th#1\not$%
    }%
    {#2}%
  }%
}
\makeatother

\begin{document}

\title{Insights into the ``cross-world'' independence assumption of causal mediation analysis}
\author{Ryan M. Andrews; Vanessa Didelez}
\date{}
\maketitle

\begin{abstract}
\noindent Causal mediation analysis is a useful tool for epidemiological research, but it has been criticized for relying on a ``cross-world'' independence assumption that is empirically difficult to verify and problematic to justify based on background knowledge. In the present article, we aim to assist the applied researcher in understanding this assumption. Synthesizing what is known about the cross-world independence assumption, we discuss the relationship between assumptions for causal mediation analyses, causal models, and non-parametric identification of natural direct and indirect effects. In particular, we give a practical example of an applied setting where the cross-world independence assumption is violated even without any post-treatment confounding. Further, we review possible alternatives to the cross-world independence assumption, including the use of bounds that avoid the assumption altogether. Finally, we carry out a numerical study in which the cross-world independence assumption is violated to assess the ensuing bias in estimating natural direct and indirect effects. We conclude with recommendations for carrying out causal mediation analyses. 
\end{abstract}


\newpage

\section*{Introduction}

Causal mediation analysis provides a formal framework for investigating causal pathways between an exposure (equivalently, treatment) and an outcome of interest \cite{robinsgreenland1992, pearl2001}. As with all causal methods, the identification of causal mediation effects from observed data require that certain structural assumptions hold. Violations of these assumptions, like positivity and consistency, can be subtle, and there is sometimes no universal agreement over what exactly the assumptions mean in practice \cite{westreich2010positivity, cole2009consistency, vanderweele2009consistency, pearl2010consistency}. With respect to causal mediation analysis, there have been controversies over the so-called ``cross-world'' independence assumption \cite{richardson2013swig, naimi2014mediation, robinsrichardson2011} and the public health relevance of the estimands known as natural direct effects (NDE) and natural indirect effects (NIE) due to this assumption \cite{naimi2014mediation,kaufman2010invited, robinsrichardson2011, naimi2015boundless}. Some of these arguments have even prompted Robins and Richardson \cite{richardson2013swig} to develop ``single-world intervention graphs'' (SWIGs) which help the user to entirely avoid any assumptions across different worlds in causal analyses. Relatedly, others have developed different estimands for causal mediation, like ``standardized'', ``interventional'' or ``separable'' effects, that avoid cross-world notions and rely on other structural assumptions about how the exposure, mediator, and outcome relate to each other \cite{geneletti2007identifying, robinsrichardson2011, lok2016defining, moreno2018understanding, vansteelandt2017interventional, didelez2019defining, robins2020interventionist}.  

\bigskip

In the language of counterfactuals, the cross-world independence assumption is that

\begin{equation}
Y(a,m) \independent M(a') \qquad \forall m \label{CW}
\end{equation}

where the counterfactual $Y(a,m)$ is the value of the outcome $Y$ that would be observed if, possibly counter to fact, exposure $A$ were set to $A=a$ and the mediator $M$ were set to $m$, and where $M(a')$ is the value of $M$ under the assignment $A=a'$, with possibly $a' \ne a$ (for our purposes, we assume that $a \neq a'$). 
In words, this assumption is that there is an independence between counterfactual outcome and mediator values ``across worlds," with one being a world in which the exposure is set to $A=a$ for the outcome and the other being a world in which it is set to $A=a'$ for the mediator. Such an exposure assignment cannot occur in real-life, making the cross-world independence assumption impossible to verify, even in principle, without relying on other equally problematic assumptions.

\bigskip

Most of what is known about the cross-world assumption appears scattered across technical journals spanning multiple disciplines. Furthermore, it is difficult to translate the precise mathematical statement of the cross-world independence assumption in \eqref{CW} into practical examples or words, which has led to some incomplete descriptions of it, particularly with respect to its role in the non-parametric identification of causal mediation effects. In this paper, we provide a comprehensive overview of the cross-world independence assumption by focusing on two particular points that deserve greater clarification. 
First, it is usually equated with the absence of any ``intermediate confounding,'' i.e., confounding between $M$ and $Y$ by a variable that is itself affected by exposure \cite{pearl2001, destavola2014sem, vanderweele2014effect,avin2005identifiability}. However,  this presupposes that the data were generated by a non-parametric structural equation model with independent errors (NPSEM-IE) \cite{robinsrichardson2011}.
A second sometimes overlooked issue is that the cross-world independence assumption is not a  necessary condition for the identifiability of causal mediation effects \cite{naimi2014mediation, keele2015causal, vanderweele2016mediation, destavola2014sem}. We briefly discuss alternative estimands for causal mediation  which redefine the target of inference and thus somewhat avoid assumption \eqref{CW}. We conclude with a numerical illustration to investigate the amount of bias one could face in a hypothetical (but not implausible) scenario where specifically assumption \eqref{CW} and no other assumption of causal mediation is violated.

\section*{Violations of the cross-world independence assumption are not limited to intermediate confounding}

The natural direct and indirect effects, NDE and NIE, are defined as the counterfactual contrasts \cite{robinsgreenland1992,pearl2001}

\begin{equation}
\text{NDE} = \mathbb{E}\left\{ Y\left(a, M(a')\right) - Y\left(a',M(a')\right) \right\} \label{NDEtraditional}
\end{equation}
and
\begin{equation}
\text{NIE} = \mathbb{E}\left\{ Y\left(a, M(a)\right) - Y\left(a,M(a')\right) \right\} \label{NIEtraditional}
\end{equation}

In words, the NDE is the average effect on $Y$ of setting the exposure to $A=a$ versus $A=a'$, but for each individual $M$ is set to the value it would have taken under $A=a'$. The NIE is the average effect on $Y$ if $A$ were fixed at $A=a$, but the mediator $M$ was set to the value it would take under $A=a$ versus $A=a'$. To establish non-parametric identifiability based on a given causal model, one typically shows that (\ref{NDEtraditional}) and (\ref{NIEtraditional}) correspond to an observable data contrast under the restrictions imposed by the model \cite{pearl2001, pearl2014interpretation, avin2005identifiability, shpitser2013longmed}. 
Note that, throughout, we take it for granted that positivity  \cite{westreich2010positivity} and consistency \cite{cole2009consistency, vanderweele2009consistency} hold.

\begin{figure}[ht]
\begin{center}
\begin{tikzpicture}
\tikzstyle{format} = [draw, very thick, circle, minimum size=1.25cm, inner sep=0pt]

\node[format](A){$A$};
\node[format, right=2cm of A](M){$M$};
\node[format, right=2cm of M](Y){$Y$};

\draw[very thick,->](A) to (M);
\draw[very thick,->](M) to (Y);
\draw[very thick, bend left, ->, dashed](A) to (Y);

\end{tikzpicture}
\end{center}
\caption{Example of a DAG representing mediation of the effect of the exposure $A$ on the outcome $Y$ by the mediator $M$.}
\label{simpledag}
\end{figure}
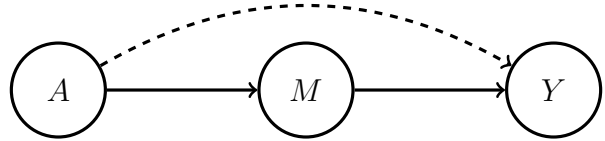

For example, suppose that we have data on $A$, $M$, and $Y$, and that Figure \ref{simpledag} is the causal directed acyclic graph (DAG) which we believe generated the data (note that here we assume no measured or unmeasured confounding between variables). This is in fact not specific enough to non-parametrically identify the NDE or NIE, as the DAG could either represent (a) Robins' Finest Fully Randomized Causally Interpretable Structured Tree Graph (FFRCISTG) model \cite{robins1986} or (b) Pearl's NPSEM-IE \cite{pearl2001}. Let Figure \ref{simpledag} represent a FFRCISTG model, which then implies the following restrictions on the counterfactuals $M(a)$ and $Y(a,m)$:
\begin{eqnarray}
M(a) &\independent& A \label{finest1} \\
Y(a,m) &\independent& A \label{finest2}  \\
Y(a,m) &\independent& M(a) \mid A=a \label{finest3}   ,
\end{eqnarray}
for all $a$ and $m$. 
While (\ref{finest1}) and (\ref{finest2}) can be ensured by randomizing $A$, (\ref{finest3}) often only holds under additional conditioning on a suitable set of covariates to account for confounding between $M$ and $Y$ other than by $A$; we omit this here given our simple example with no confounding.

\bigskip

A key feature of the FFRCISTG model is that it is a ``single-world" model \cite{richardson2013swig}:  it does not allow us to make statements about counterfactuals under different interventions on the same variable. Thus, \eqref{finest3} has the same value $a$ in all three places, and all three assumptions can be checked on a SWIG. Robins and Greenland \cite{robinsgreenland1992} discuss assumptions that would allow non-parametric identification of the pure or total direct and total or pure indirect effect (i.e., the NDE and NIE) under a FFRCISTG model, and conclude that assumptions \eqref{finest1} - \eqref{finest3} are insufficient. They  point out that \eqref{CW} should hold for the NDE and NIE to be non-parametrically identified, but that checking this assumption would require a ``crossover trial without carryover effects"  in which an individual is observed under the exposed condition, all conditions are returned to what they were under the pre-exposed state, and then the individual is observed under the unexposed condition (or vice-versa). Such a trial is rarely possible, and so they claim that alternative assumptions are needed to identify the NDE and NIE under a FFRCISTG model, as will be discussed in the next section.

\bigskip

In contrast, Pearl \cite{pearl2001} shows that the NDE and NIE are non-parametrically identified under a NPSEM-IE and causal DAG of Figure \ref{simpledag}. This is because an NPSEM-IE is a more restrictive model class than a FFRCISTG, imposing additional independence conditions across ``multiple-worlds" which cannot be represented in a SWIG; if Figure \ref{simpledag}  instead represents an NPSEM-IE, not only would (\ref{finest1}) - (\ref{finest3}) be implied, but also assumption \eqref{CW}, i.e., under a NPSEM-IE with the DAG of Figure \ref{simpledag}, $Y(a,m)$ and $M(a')$ are independent even if $a \neq a'$. Under a FFRCISTG model, this independence only holds for $a=a'$, where this value must also agree with the observed $A=a$, cf. (\ref{finest3}). Because assumption \eqref{CW} imposed by the NPSEM-IE is so powerful that it yields identification of NDE and NIE, it should be well-understood and carefully assessed. 

\bigskip

When assumptions (\ref{CW}) and (\ref{finest1}) - (\ref{finest3}) hold, we have that $\mathbb{E}\{Y(a,M(a'))\}$ is identified by the so-called mediational g-formula
$$
\mathbb{E}\left\{Y(a,M(a'))\right\}  = \sum_m \mathbb{E}\left\{Y \mid A=a, M=m \right\}p(M=m \mid A=a'). 
$$
where the sum is replaced by an integral in the case of a continuous mediator \cite{pearl2001}. This results in the following observed data contrasts for the NDE and NIE:
\begin{eqnarray}
\widehat{NDE} &=& \sum_m \left( \mathbb{E}\left\{Y \mid A=a, M=m \right\} - \mathbb{E}\left\{ Y \mid A=a', M=m \right\} \right)p(M=m \mid A=a') \label{NDEformula}\\
\widehat{NIE} &=& \sum_m \mathbb{E}\left\{Y \mid A=a, M=m \right\}\left( p(M=m \mid A=a) - p(M=m \mid A=a')\right) \label{NIEformula}
\end{eqnarray}
When some of the assumptions are violated, we may still compute or estimate $\widehat{NDE}$ and $\widehat{NIE}$, but they are then possibly biased for the true NDE and NIE, respectively. 

\bigskip

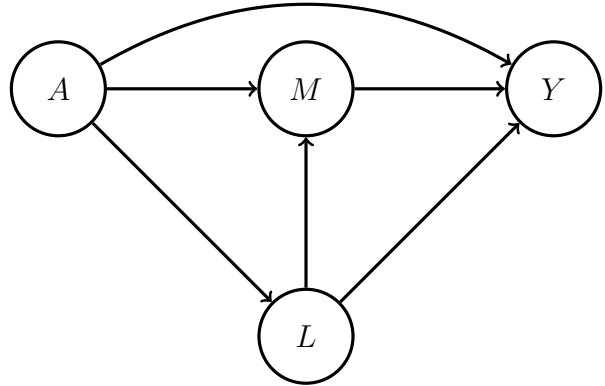
\begin{figure}[ht]
\begin{center}
\begin{tikzpicture}
\tikzstyle{format} = [draw, very thick, circle, minimum size=1.25cm, inner sep=0pt]

\node[format](A){$A$};
\node[format, right=2cm of A](M){$M$};
\node[format, below=2cm of M](L){$L$};
\node[format, right=2cm of M](Y){$Y$};

\draw[very thick, ->](A) to (M);
\draw[very thick, ->](M) to (Y);
\draw[very thick, ->](L) to (Y);
\draw[very thick, bend left, ->](A) to (Y);
\draw[very thick, ->](A) to (L);
\draw[very thick, ->](L) to (M);

\end{tikzpicture}
\end{center}
\caption{Example of a DAG representing mediation of the effect of the exposure $A$ on the outcome $Y$ by the mediator $M$, but there is another variable $L$ that is a mediator-outcome confounder affected by exposure. The presence of this "intermediate confounder" means the NDE and NIE of $A$ on $Y$ through $M$ are not non-parametrically identified.}
\label{fig:recanting}
\end{figure}

All epidemiologic studies that adopt the assumptions of Pearl (2001)\cite{pearl2001} as part of their causal mediation analysis assume, either implicitly or explicitly, that their data are generated by a NPSEM-IE. Through DAGs, the causal mediation assumptions are generally framed within graphical criteria given by Pearl for \eqref{CW} and \eqref{finest1} - \eqref{finest3} to be satisfied, one of which is that there is no mediator-outcome confounder affected by exposure, like the variable $L$ in Figure \ref{fig:recanting}, as this violates \eqref{CW} {\it even when data on $L$ is available}. Consequently, violations of \eqref{CW} tend to be equated with the presence of such intermediate confounding  \cite{destavola2014sem,miles2017quantifying, rudolph2019causal}.  Pearl's graphical criteria have been extended into a general theory of non-parametric identification of path-specific effects. It turns out that under a NPSEM-IE, a violation of \eqref{CW} alone essentially \textit{only} occurs when there is (observed) intermediate confounding while (\ref{finest1})  and (\ref{finest2}) hold, and (\ref{finest3}) holds conditionally on $L$ (a more precise statement requires the notion of ``recanting witness''\cite{avin2005identifiability, shpitser2013longmed}).

\bigskip

To give an idea why the NDE and NIE are not generally identified in the presence of an $L$, even if observed, as in Figure \ref{fig:recanting}, we consider the corresponding NPSEM-IE:
\begin{eqnarray*}
A &=& f_A(\epsilon_A)\\
L &=& f_L (A, \epsilon_L)\\
M &=& f_M(A, L, \epsilon_M)\\ 
Y &=& f_Y(A, L, M, \epsilon_Y),
\end{eqnarray*}
where the $\epsilon$ error terms represent both random and systematic variation (i.e., variation occurring from omitted or latent variables). By definition, $\epsilon_A \independent \epsilon_L \independent \epsilon_M \independent \epsilon_Y$. This system of equations implies the following construction of counterfactuals:
\begin{eqnarray*}
L(a) &=& f_L(a, \epsilon_L)\\
L(a') &=& f_L(a', \epsilon_L)\\
M(a') &=& f_M(a', L(a'), \epsilon_M)\\
Y(a,m) &=& f_Y(a, m, L(a), \epsilon_Y). 
\end{eqnarray*}

Importantly, the error terms for the factual variables are assumed to be exactly the same for their corresponding counterfactuals. For example, $\epsilon_L$ is assumed to be the same for $L$, $L(a)$, and $L(a')$. Due to the independence of the different error terms, it is straightforward to confirm that the assumptions (\ref{finest1}) and (\ref{finest2}) hold in this system, and (\ref{finest3}) holds upon additional conditioning on $L$. However, (\ref{CW}) does not hold, with or without conditioning on $L$, because  $L(a) \not\independent L(a')$ in general due to the common error term. Therefore, the NDE and NIE cannot be non-parametrically identified in a NPSEM-IE when there is an intermediate confounder $L$, regardless of whether it is observed or unobserved \cite{pearl2001, avin2005identifiability, shpitser2013longmed, steen2018mediation}.

\bigskip

We now turn to the question of how the cross-world independence \eqref{CW} can be violated other than by intermediate confounding even in cases where assumptions (\ref{finest1}) - (\ref{finest3}) hold. Robins and Richardson \cite{robinsrichardson2011} give an example where \eqref{CW} does not hold for a data generating mechanism that is not an NPSEM-IE and exhibits no intermediate confounding. Consider Figure \ref{RobinsRichardsonFig11}, which is similar to Figure \ref{simpledag} but counterfactuals for $M$ and $Y$ have been added along with an unobserved (i.e., latent) variable $U$ that affects specifically $M(0)$ and $Y(1,m)$. For simplicity, we consider a binary exposure with exposed being defined as $A=1$ and unexposed as $A=0$; however, the argument is similar for the more general $A=a$ versus $A=a'$ case. The structural equations corresponding to Figure \ref{RobinsRichardsonFig11} are

\begin{eqnarray*}
U  &=& f_U(\epsilon_U) \\ \nonumber
A &=& f_A(\epsilon_A) \nonumber
\end{eqnarray*}

\begin{equation}
M= f_M(A, U, \epsilon_M) = 
\begin{cases}
  \tilde{f}_{1, M}(\epsilon_M), & \text{if } A=1\\

  \tilde{f}_{0, M}(U, \epsilon_M), & \text{if } A=0 \nonumber
\end{cases}
\end{equation}

\begin{equation}
Y= f_Y(A, M, U, \epsilon_Y) = 
\begin{cases}
  \tilde{f}_{1, Y}(M,U,\epsilon_Y), & \text{if } A=1\\

  \tilde{f}_{0, Y}(M,\epsilon_Y), & \text{if } A=0 \nonumber
\end{cases},
\end{equation}

where we introduced separate functions $\tilde{f}$ for each value of $A$ to make explicit that $\tilde{f}_{1, M}(\cdot), \tilde{f}_{0, Y}(\cdot)$ are not functions of $U$. 
As before, the $\epsilon$ error terms of this system of structural equations are considered independent of each other. However, if $U$ is unobserved, the structural equation model on $(A, M, Y)$ alone does not have independent error terms. Instead, the error term for $M$, say $\tilde{\epsilon}_M$,  is a combination of $\epsilon_M$ and $U$, since the unobserved $U$ becomes part of the error term. Similarly, the error term for $Y$, say $\tilde{\epsilon}_Y$, is a combination of $\epsilon_Y$ and $U$. This means that $\tilde{\epsilon}_M \not\independent \tilde{\epsilon}_Y$ and the system of equations on the observable $(A,M,Y)$ is not an NPSEM-IE. This has implications for the counterfactuals; we have that

$$M(0) = \tilde{f}_{0,M}(\tilde \epsilon_M) $$
$$M(1) = \tilde{f}_{1,M}(\epsilon_M)$$
$$Y(0,m) = \tilde{f}_{0,Y}(m, \epsilon_Y) $$
$$ Y(1, m) = \tilde{f}_{1,Y}(m, \tilde \epsilon_Y).$$

Thus, the cross-world assumption (\ref{CW}) is violated as $Y(1,m) \not\independent M(0)$  due to the lack of independence between $\tilde{\epsilon}_M$ and $\tilde{\epsilon}_Y$.
Crucially, however, $U$ is {\em not} an ordinary confounder of $M$ and $Y$ because assumption \eqref{finest3}  still holds (as do assumptions \eqref{finest1} and \eqref{finest2}), as is easily checked. In other words, under the model assumptions of the above example, we can identify the total causal effect of $A$ on $M$ or on $Y$, and the causal effect of $M$ on $Y$. We will call $U$ a ``cross-world confounder.''

\bigskip

The example demonstrates the special role of the cross-world independence assumption \eqref{CW}; it is therefore important to justify it separately from any of the other assumptions. One should not only assess whether it may be violated through intermediate confounding, but also whether it is plausible that one's data was generated by a NPSEM-IE, in particular whether there are latent variables acting as ``confounders'' across the different interventional settings of the exposure $A$.

\subsection*{\textit{Illustration: Knee surgery and quality of life}}

While there are many plausible practical examples for intermediate confounding \cite{destavola2014sem,miles2017quantifying, rudolph2019causal}, violations of assumption
\eqref{CW} without intermediate confounding (i.e., not assuming a NPSEM-IE) in real applications are harder to imagine due to the cross-world aspect. As mentioned earlier, empirical evidence would need to come from a kind of cross-over trial; if this is not available, detailed subject matter understanding of the biological mechanisms involved is required. Here, we provide a possible real-world illustration of the kind of reasoning needed to pinpoint cross-world confounding.

\bigskip

Suppose an investigator is interested in estimating both the direct and indirect effect of knee replacement surgery on quality of life among older adults over 50 years of age through walking speed. Numerous studies have found that undergoing this surgery significantly improves both walking speed \cite{weidenhielm1993improvement, mattsson1990assessment,berghmans2018functions} and quality of life \cite{brandes2011changes,dailiana2015patient}, and walking speed has been shown to be an important factor for quality of life as well \cite{ekstrom2011effects,sarkisian2007pilot}; therefore, the hypothesis that walking speed may mediate the effect of the surgery on quality of life is plausible. Define $A=1$ to be knee replacement surgery and $A=0$ to be no surgery. Let $M$ denote whether walking speed is normal versus slow, with normal walking speeds exceeding 0.8 m/s ($M=1$) and slow walking speeds falling in the range of $\le 0.8$ m/s ($M=0$). Further, let $Y$ denote quality of life, measured on a continuous scale. The investigator would therefore like to estimate the NDE and NIE of $A$ on $Y$ through $M$.

\bigskip

In addition, suppose (possibly unknown to the investigator) that there exists a genetic marker $U$ with the following properties:

\begin{enumerate}
\item Its effect starts to occur around the age of 50, so $U$ is unlikely to affect any possible exclusion criteria like a slow walking speed at baseline.
\item When it is present, it adversely affects knee joints and causes individuals to have a slower walking speed than if it were absent, meaning that individuals are less likely to have a walking speed $\ge 0.8$ m/s. Knee replacement surgery completely removes the influence of $U$ on walking speed, however, because it can no longer affect the (now artificial) knee joint.
\item When it is present, individuals over 50 have a higher probability of surgical scarring since $U$ interferes with the body's ability to heal. This negatively impacts quality of life.
\item $U$ cannot affect quality of life except through its relationships with walking speed and surgical scarring.
\end{enumerate}

While no study has discovered a genetic marker with exactly all of these properties, prior human and animal studies have found that walking speed in later life has a genetic component \cite{tiainen2007genetic,pajala2005contribution, ortega2006twin, lunetta2007genetic}, that scarring has a genetic component \cite{gallant2006genetic, kachgal2012dual, gallant2006skin,van2009potential, malfait2005molecular}, and that scarring affects quality of life \cite{bock2006quality,brown2008hidden}. Therefore, we believe it is reasonable to assume that such a $U$ could exist. For our example, we further assume that surgical scarring was unmeasured, so it is treated as a latent variable.

\bigskip

The causal structure of the above example can be represented by the DAG of Figure \ref{RobinsRichardsonFig11} on $(A,M,Y,U)$, but in order to make the role of the latent variable ``scarring'' on $Y$ and $M$ explicit, we have expanded the DAG by a node $S$ as given in Figure \ref{KneeIntermediateC}. 
In Figure \ref{RobinsRichardsonFig11}, the $A \rightarrow S \rightarrow Y$ pathway would be considered part of the  $A \rightarrow Y$ edge. However, with Figure \ref{KneeIntermediateC} we can make explicit the additional crucial assumption that scarring does not affect walking speed by the absence of the $S\rightarrow M$ edge (shown dotted). This assumption is reasonable if scarring is never severe enough to physically impede the knee joint, or if knee scarring  never causes one's walking speed to fall below $0.8$ m/s. Hence, in this case we have that assumptions (\ref{finest1}) - (\ref{finest3}) hold. However, the cross-world assumption  \eqref{CW}  is still violated because $Y(1,m) \not\independent M(0)$ due to the cross-world confounder $U$ acting on $Y(1,m)=Y(1,S(1),m)$ via $S(1)$.
While neither $S$ nor any other variables induce intermediate confounding, the investigator would thus still be mistaken in believing that the NDE and NIE are non-parametrically identified if he or she used the absence of intermediate confounding as the sole criterion for justifying the cross-world assumption.\\
Note that another researcher might be more comfortable assuming the presence of an $S\rightarrow M$ edge in which case $S$ becomes an (unobserved) intermediate confounder, so that identification does not hold for more reasons than violation of \eqref{CW} by $U$: it will also be violated by $S$, and (\ref{finest3}) is violated as we cannot adjust for the latent $S$.
These kind of considerations underscore the importance of subject-matter knowledge in causal analyses to ensure that the DAG one works with is indeed a causally sufficient DAG that captures all relevant variables and relationships. Without this knowledge, it is difficult to assess how plausible one's causal assumptions are, including the cross-world independence assumption. 

\section*{Alternatives to the cross-world assumption for identification of causal mediation effects}

Because Pearl's seminal paper shows that the NDE and NIE are non-parametrically identified under assumptions including cross-world independence \cite{pearl2001}, and because Robins and Greenland argue that these effects are not identified under any causal model that does not include any cross-world independence assumptions (like an FFRCISTG model) \cite{robinsgreenland1992}, it is possible to mistakenly assume that \eqref{CW} must be a necessary assumption for the NDE and NIE to be identified. 

\bigskip

However, the cross-world independence assumption (together with (\ref{finest1}) - (\ref{finest3})) is a sufficient, but not a necessary, assumption \cite{pearl2014interpretation, tchetgen2014identification, imai2010identification, petersen2006estimation, hafeman2011alternative}. 
While assumptions \eqref{finest1} - \eqref{finest3} \textit{alone} cannot identify the NDE and NIE, alternatives approaches are possible as we discuss next.

\subsection*{\textit{Bounds and sensitivity analysis}}

Under assumptions  \eqref{finest1} - \eqref{finest3} alone, the NDE and NIE are not point-identified, i.e., even with an infinitely large sample they cannot be narrowed down to a single value. 
However, in certain settings one can derive bounded effects \cite{cai2008bounds,kaufman2005improved,robinsrichardson2011}.
Care must be taken with the interpretation of such bounds, as these are not to be confused with confidence intervals. The bounds give the whole mathematically possible range of values for the NDE (or NIE) that are compatible with the observed data on $(A,M,Y)$ under the chosen model assumptions. 

\bigskip

For instance, assuming only \eqref{finest1} - \eqref{finest3} and when all variables, $A,M,Y$, are binary, the bounds for the NDE are given as follows \cite{robinsrichardson2011} (with $a=1, a'=0$):

\begin{multline}
\max \left(0, p(M=0 \mid A=0) + \mathbb{E}\{Y \mid A=1, M=0\} -1 \right) + \\ \max \left(0,p(M=1 \mid A=0) + \mathbb{E}\{Y \mid A=1, M=1\} -1  \right) - \mathbb{E}\{Y \mid A=0\} \\ \le NDE \le \\ \min \left(p(M=0 \mid A=0), \mathbb{E}\{Y \mid A=1, M=0\} \right) + \\ \min \left( p(M=1 \mid A=0), \mathbb{E}\{Y \mid A=1, M=1\}\right) - \mathbb{E}\{Y \mid A=0\} \label{NDEbounds}
\end{multline}

These bounds are valid under arbitrary violations of \eqref{CW} as they do not rely on  cross-world independence, and they have been extended to cases where there is possible intermediate confounding \cite{tchetgen2014bounds} and when the mediator is polytomous \cite{miles2015partial}. Unless the bounds are from $-1$ to $1$, they are ``informative'' in the sense that they exclude some impossible values for the NDE. However, when the bounds are wide and contain zero, they are often regarded as uninformative, and some authors have pointed out that sensitivity analyses exploiting prior knowledge to limit how strongly an assumption is violated is more useful in practice \cite{jiangvanderweele2015, imai2010identification}. Nevertheless, it is good practice to report the bounds when possible, since they quantify what can be inferred from the observable data together with the assumptions \eqref{finest1} - \eqref{finest3} but without the cross-world independence assumption, without parametric model assumptions, and without any prior knowledge. If they are wide, it means that any narrowing down by imposing \eqref{CW} or other restrictions, as discussed below, crucially hinges on those additional assumptions and restrictions, so that these must be carefully discussed in any given application.

\bigskip

Sensitivity analysis techniques for causal effects allow an analyst to specify varying levels of unobserved confounding between variables, and then for each level, assess how the causal effect estimate of interest would change had this unobserved confounding been removed (e.g., by adjustment) \cite{lash2011applying}. With respect to the NDE and NIE, available sensitivity analyses primarily focus on unobserved mediator-outcome confounding, either by specifying parameters quantifying the unobserved confounding itself \cite{leCessie2016bias} or by specifying a correlation between the error terms of the structural models for $M$ and $Y$ \cite{imai2010identification, albert2014sensitivity, lindmark2018sensitivity}. Some of these methods are specifically designed for assessing violations of \eqref{CW} due to intermediate confounding \cite{tchetgen2012semiparametric, hong2018weighting, vanderweele2014sensitivity, imai2013identification, vansteelandt2012natural}. However, to our knowledge, no sensitivity analysis techniques have been developed for the case of cross-world confounding, like in Figure \ref{RobinsRichardsonFig11}.

\subsection*{\textit{Parametric assumptions}}

Causal mediation analyses are often based on specific models, where the parametric assumptions allow for the identification of the NDE and NIE even under violations of \eqref{CW}. 
For example, in the presence of intermediate confounding by $L$ (Figure \ref{fig:recanting}), one could assume a linear structural equation model with independent errors (LSEM-IE) for the relationships between $A$, $L$, $M$, and $Y$:
\begin{align}
L &= \alpha_A A + \epsilon_L  \nonumber  \\
M &= \beta_A A + \beta_L L + \epsilon_M   \nonumber  \\
Y &= \theta_A A + \theta_L L +  \theta_M M  + \epsilon_Y \nonumber 
\end{align}

Earlier, we showed that the NDE and NIE of $A$ on $Y$ through $M$ were not non-parametrically identified in general due to $L$. However, by making the above linearity assumption, the contrast $L(a) - L(a') = \alpha_A(a - a')$ is a constant individual-level effect, which allows for the NDE and NIE to be identified: 
\begin{align}
Y(a, M(a')) &= Y(a, L(a), M(a', L(a'))) \nonumber \\
&= \theta_A a + \theta_L(\alpha_A a + \epsilon_L) + \theta_M(\beta_A a' + \beta_L(\alpha_A a' + \epsilon_L) + \epsilon_M) + \epsilon_Y \nonumber \\
&= \left( \theta_A + \theta_L \alpha_A \right)a + \left( \theta_M \beta_A + \theta_M \beta_L \alpha_A \right)a' + \left( \theta_L + \theta_M\beta_L\right)\epsilon_L + \theta_M\epsilon_M + \epsilon_Y \label{eq:intermconf}
\end{align}

Here, the NDE is the term in front of $a$, while the NIE is the term in front of $a'$. Interestingly, $\alpha_A$ appears in both the NDE and NIE, owing to the fact that pathways through $L$ belong to both the direct and indirect effect.

\subsection*{\textit{Relaxing linearity}}

One can somewhat relax the above simple LSEM  and still obtain identification of the NDE and NIE (for more complex models incorporating higher-order terms and interactions  (see De Stavola et al. \cite{destavola2014sem}).
Robins and Greenland \cite{robinsgreenland1992} also show that the NDE and NIE are identified under a FFRCISTG model when there is no individual-level additive interaction, i.e., when 
\begin{equation} 
Y(a,m) - Y(a',m)=B(a,a'),  \label{eq:Baa}
\end{equation}  
where $B(a,a')$ is a random function that does not depend on $m$. This is because under this assumption, the NDE is simply $\mathbb{E}\{B(a,a')\}$, and this function is identifiable under the FFRCISTG assumptions. Therefore, if an investigator can argue on substantive grounds that the difference in \eqref{eq:Baa} does not vary across levels of the mediator, the non-parametric identification of the NDE and NIE is possible. A similar approach is given by Petersen et al. \cite{petersen2006estimation}, who show that the NDE can be identified if (\ref{CW}) is replaced with a ``direct effect assumption:"
\begin{equation} 
\mathbb{E}\left\{Y(a,m) - Y(a',m) \mid M(a')=m\right\} = \mathbb{E}\left\{Y(a,m) - Y(a',m)\right\} \label{DEassump}
\end{equation}

It is easy to see that (\ref{eq:Baa}) implies (\ref{DEassump}), but the latter is slightly weaker. Moreover, (\ref{DEassump}) is implied by (\ref{CW}) but there are situations where (\ref{DEassump}) holds but (\ref{CW}) does not  \cite{van2004estimation, petersen2006estimation}. Finally, with respect to intermediate confounding by $L$ in particular (Figure \ref{fig:recanting}), Robins and Richardson \cite{robinsrichardson2011} also point out that the NDE and NIE is identified if $L(a)$ is a deterministic (not necessarily linear) function of $L(a')$, since this allows one to substitute $g(L(a'))$ for $L(a)$ for some function $g(\cdot)$ when identifying $\mathbb{E}\{Y(a,M(a')\}$, generalizing (\ref{eq:intermconf}). Such a function is guaranteed to exist, and can in principle be estimated from observable data, for a continuous scalar $L$ under the condition of rank preservation, i.e.,

$$L_i(a') < L_j (a') \Rightarrow L_i(a) < L_j(a)$$

for all individuals $i$ and $j$ (e.g., the quantile-quantile function). Note that out of the above parametric assumptions, LSEM-IE is the strongest, as it implies both no individual-level additive interaction and rank preservation by definition. 
This last extension does not appear to have been used in any practical application yet.

\bigskip

Finally, we  point out that the complete characterization for when the NDE and NIE (and path-specific effects more generally) are non-parametrically identified has been given by Shpitser (2013) \cite{shpitser2013longmed}. In short, there exist specific graphical structures with which it is still possible to non-parametrically identify the NDE and NIE even if \eqref{CW} is violated. For example, if there is a variable that itself fully mediates the $A-Y$ or $A-M$ relationship, this ``mediating instrument'' can be used to non-parametrically identify the NDE and NIE without \eqref{CW}. We refer the reader to this paper, along with several others\cite{pearl2014interpretation,  steen2018mediation}, for the technical details and examples.

\section*{Alternative estimands}

Over the last decade, alternative approaches to causal mediation focusing on different estimands have been proposed. These have sometimes been motivated by a practical impossibility to intervene on and set the mediator $M$ to a given value, let alone to fix it at $M(a')$, or a desire to avoid the ``cross-world'' notion of the nested counterfactual $Y(a,M(a'))$ itself and assumption \eqref{CW}. For example, so-called ``standardized'', ``interventional'' or ``organic'' concepts represent the effect of an intervention on the mediator which generates $M$ randomly from a conditional distribution given an exposure assignment, rather than fixing the mediator to an individual-specific value \cite{geneletti2007identifying, vanderweele2014effect,  lok2016defining}. Another alternative considers ``separable effects," which assumes that $A$ consists of two components $A^M$ and $A^Y$ that affect the outcome only through the mediator or only \textit{not} through the mediator, respectively\cite{robinsrichardson2011,didelez2019defining,aalen2019time,stensrudJASA}. Observationally $A=A^M = A^Y$,  but separability means that one can conceive of  $A^M$ and  $A^Y$ as separately intervenable in a single world so that the counterfactual $Y(A^Y=a, A^M=a')$ replaces $Y(a,M(a'))$; this also shifts  the focus away from interventions on $M$, and instead to interventions on different aspects of the exposure $A$.  \\

The above approaches make do without cross-world notions of effects or assumptions and are thus, at least in principle, falsifiable by suitable experiments. However this comes  at the ``cost'' of new estimands redefining direct and indirect effects, which requires care in interpretation and can also lead to some counter-intuitive results. For example, ``interventional" direct and indirect effects may not sum to the total effect of $A$ on $Y$, particularly in the presence of intermediate confounding \cite{didelez2006direct,vansteelandt2017interventional}. In the Supplementary Materials, we provide greater detail on these alternative approaches (see also recent discussions\cite{rudolph2019causal, nguyen2019clarifying, robins2020interventionist}). 
We see the value of the alternative causal mediation estimands in motivating the analyst to think carefully about the desired target of inference in a practical application, and to look at the structural assumptions from different angles.

\section*{Numerical illustration}

While bias due to intermediate confounding has been the subject of several sensitivity analysis papers, no attention has been given to investigating the bias that could arise from violations due to cross-world confounding like in Figure \ref{RobinsRichardsonFig11}. Therefore, we designed a study to investigate this bias based on a range of settings following the example of knee replacement surgery, walking speed, and quality of life described earlier. Specifically, we evaluated the discrepancy  between 
the true values of the NDE and NIE and Pearl's mediation formula estimands (i.e., \eqref{NDEformula} and \eqref{NIEformula}). We deliberately designed our data generating models to not have any intermediate confounding and to not satisfy any alternative assumptions that would possibly allow for identification without the cross-world assumption \cite{petersen2006estimation,hafeman2011alternative}.

\bigskip

For convenience, we assumed that $M$ is a binary variable, but considered both binary and continuous $Y$. We assumed that Figure \ref{RobinsRichardsonFig11} represented the true causal DAG, that $A$ is a flip of a fair coin, that $U$ is drawn from a $ \text{Normal}(\mu=2, \sigma=1)$ distribution, and the structural models for $M$ and $Y$ were defined to be

\begin{equation}
f_M(A, U, \epsilon_M)=
\begin{cases}
1, & \text{if} \qquad -\epsilon_M < \alpha_0 + \alpha_1 A + \alpha_2(1-A) U \\
0, & \text{otherwise} \nonumber
\end{cases}
\end{equation}

$$f_Y(A,M,U,\epsilon_Y) = \beta_0 + \beta_1 A + \beta_2 M + \beta_3A U + \beta_4 AM + \beta_5AMU + \epsilon_Y \qquad \text{(continuous Y)}$$

\begin{equation}
f_Y(A,M,U,\epsilon'_Y)=
\begin{cases}
1, & \text{if} \qquad -\epsilon'_Y < \beta'_0 + \beta'_1 A + \beta'_2 M + \beta'_3A U + \beta'_4 AM + \beta'_5 AMU  \\
0, & \text{otherwise} \nonumber
\end{cases} \qquad \text{(binary Y)}
\end{equation}

\bigskip

where $\epsilon_M$ and $\epsilon'_Y$ are drawn from a standard logistic distribution and $\epsilon_Y$ is drawn from a standard normal distribution. Note that this model specification leads to only $M(0)$ and $Y(1,m)$ being associated due to $U$, as desired, i.e.,  the cross-world independence is violated.    Note that for the linear model generating a continuous $Y$ we show in the Supplementary Materials that when $\beta_5=0$ the  NDE and NIE are point-identified by \eqref{NDEformula} and \eqref{NIEformula}, so this case does not yield any bias. For the illustration, we chose model parameter values that covered a wide range of possible effects, while still maintaining biological and clinical plausibility. For practical reasons, almost all coefficients took on one of four values, which consisted of the two endpoint values and two equally-spaced values in between. The exception was the interaction terms $\beta_4$, which could also take on the value of 0 to reflect no $AM$ interaction. This resulted in a total of $4^7 \times 5=327,680$ parameter settings. In the Supplementary Materials, we include more details of the exact parameter values chosen, as well as additional technical details.

\subsection*{\textit{Results}}

For the continuous $Y$ case with $\beta_5 \ne 0$, we found that  the biases for NDE and NIE
both ranged between -3.3 to 3.3, or in other words, $\pm 3$ residual standard deviations, and the worst case scenarios were those with the maximum values of the interaction term. For the binary $Y$ case, when $\beta_5 = 0$, we found that biases for NDE and NIE ranged between -0.02 to 0.02, and when $\beta_5 \ne 0$, between -0.04 to 0.04 (Figure \ref{BiasPlot}). Again, the most extreme biases were seen when the interaction terms (particularly the three-way interaction term) were at or near their maximum value.

\bigskip

To see if we could make biases even more extreme, we conducted a secondary evaluation where we allowed the parameter values to vary beyond what we believed to be biologically plausible. For example, setting the parameters to

\[
\begin{bmatrix}\alpha_0 & \alpha_1 & \alpha_2 & \beta_0  & \beta_1 & \beta_2 & \beta_3 & \beta_4 & \beta_5 \end{bmatrix} = \begin{bmatrix}-3.5 & 0.5 &  2.5  & -4 & -1  & 3.5  & 3.25 & 3 & -5 \end{bmatrix}
\]

led to a bias in NDE of $-0.18$ and in NIE of $=0.18$ for binary $Y$. We believe these parameter values are implausible for our particular knee replacement example. For instance, it would suggest that among those who do not have knee surgery, on average having a walking speed $> 0.8$ m/s vs. $\le 0.8$ m/s leads to $\exp(3.5) = 33.1$ greater odds of reporting high quality of life. With the exception of a few well-known examples (e.g., smoking on cancer risk), it is rare that any clinical or epidemiological exposure will have such an extreme impact on any outcome. 

\bigskip

For the binary $Y$ case, we also computed the bounds shown in Table \ref{tab1}. The bounds illustrate that, without assuming the cross-world independence \eqref{CW}, data generated from the above models support a wide range of possible  NDE values. While the specific data generating models formulate one way of violating  \eqref{CW}, many other ways are possible and would give rise to these different NDE values without any empirical way of narrowing these down based on the observable data.

\bigskip

\section*{Discussion}

In this paper, we aimed at providing insights into the cross-world independence assumption underlying causal mediation analysis, focusing on two issues. First, we showed that one cannot equate this assumption with the absence of an intermediate confounder: As illustrated by the knee-surgery example, cross-world confounding of $M$ and $Y$ is possible even without single-world confounding between all pairs $(A,M)$, $(A,Y)$ and $(M,Y)$ and without intermediate confounding. Second, we discussed under what alternative assumptions one can still identify the NDE and NIE even when the cross-world assumption is violated.

\bigskip

Our numerical illustration  explored the potential bias when wrongly assuming \eqref{CW}. The amount by which results ignoring cross-world confounding differed from the true NDE and NIE values depended on the settings: In the continuous $Y$ case, the most extreme difference between ${NDE}$ and the estimand $\widehat{NDE}$, over a realistic range of parameter values, was approximately 3 standard deviations. In the binary $Y$ case when the three-way interaction was non-zero, the largest difference between ${NDE}$ and $\widehat{NDE}$ was -0.04; however, the relative difference between them was almost 70\% (similar statements apply to the bias for the NIE). At the same time, the bounds for the NDE did exclude some values but were still quite wide (Table 1). All these figures are, of course, specific to our particular (and still relatively simple) data generating mechanism, but we believe they convincingly illustrate that the issue of cross-world confounding should not be ignored. 

\bigskip

We therefore recommend to anyone who wants to perform a causal mediation analysis to be mindful of the underlying assumptions, especially the cross-world independence assumption \eqref{CW}. Criticisms  in the literature of causal mediation analysis have focused on the cross-world independence assumption because it is empirically untestable, even in principle. Moreover, concepts of NDE and NIE violate the principle of ``no causation without manipulation" because there is no well-defined intervention yielding $Y(a, M(a'))$ \cite{naimi2014mediation, naimi2015boundless, robinsrichardson2011, hernan2008}.  These are valid points, but we believe that the study of causal pathways via causal mediation analysis can be both interesting and useful from a public health standpoint. We encourage researchers to use the same type of reasoning as we did in our knee surgery example and numerical illustration to decide whether the cross-world independence assumption is plausible or not for their particular application. When it is questionable, one can make additional causal assumptions, conduct a mediation analysis targeting alternative estimands \cite{geneletti2007identifying, lok2016defining,vansteelandt2017interventional, didelez2019defining, robinsrichardson2011, aalen2019time,stensrudJASA}, or utilize a number of sensitivity analysis tools to explore the possible impact of a cross-world independence violation  \cite{cai2008bounds, robinsrichardson2011, tchetgen2014bounds, hong2018weighting, daniel2015causal, vanderweele2014sensitivity, imai2013identification, vansteelandt2012natural, imai2010identification, albert2014sensitivity, lindmark2018sensitivity, vanderweele2010odds, leCessie2016bias}.

\newpage

\begin{table}
\begin{center}
 \begin{tabular}{||c c c c c||} 
\hline
\rule{0pt}{4ex}
Example & NDE & $\widehat{\text{NDE}}$ &  bias  & NDE bounds \\ [0.5ex] 
 \hline\hline
 $\beta_5 =0$  & -0.199 & -0.217 & 0.018 &  (-0.394, 0.048) \\ 
 \hline
$\beta_5 \neq 0$ & -0.055 & -0.017 & -0.037 & (-0.424, 0.566)\\[1ex] 
\hline
 extreme parameters & 0.516 & 0.616 & -0.100 & (0.141, 1.000)\\
\hline
\end{tabular}
\caption{Numerical illustration results. The first row corresponds to models in which the $AMU$ interaction $\beta_5 =0$, while the second row considers $\beta_5 \neq 0$; the "extreme parameters" row refers to the extreme setting described in the text. The bounds in each row were calculated according to Robins \& Richardson (2011), using the parameter values that resulted in the most extreme biases, and are presented as (lower, upper).}\label{tab1}

\end{center}
\end{table}

\begin{figure}[ht]
\begin{center}
\begin{tikzpicture}
\tikzstyle{format} = [draw, very thick, circle, minimum size=1.25cm, inner sep=0pt]

\node[format](A){$A$};
\node[format, right=3cm of A](M){$M$};
\node[format, below=2cm of M](Y){$Y$};
\node[format, above left=1cm of M](M1){$M(1)$};
\node[format, above right=1cm of M](M0){$M(0)$};
\node[format, above right=2.5cm of M0](U){$U$};
\node[format, below=2.5cm of U](Y11){$Y(1,1)$};
\node[format, below right=1cm of Y11](Y01){$Y(0,1)$};
\node[format, right=1cm of Y01](Y00){$Y(0,0)$};
\node[format,right=4cm of Y11](Y10){$Y(1,0)$};
\node[format, above right = 1.5cm of Y11](eY){$\epsilon_Y$};
\node[format, above =2.5 cm of M](eM){$\epsilon_M$};

\draw[very thick, line width=4pt, ->](A) to (M);
\draw[very thick, line width=4pt, ->](M) to (Y);
\draw[very thick, line width=4pt, ->](A) to (Y);
\draw[very thick,->](M0) to (M);
\draw[very thick,->](M1) to (M);
\draw[very thick,->](U) to (M0);
\draw[very thick, bend left, ->](U) to (Y10);
\draw[very thick, ->](U) to (Y11);
\draw[very thick, ->](Y11) to (Y);
\draw[very thick, bend left, ->](Y01) to (Y);
\draw[very thick, bend left=45, ->](Y00) to (Y);
\draw[very thick, bend left=70, ->](Y10) to (Y);
\draw[very thick, ->](eY) to (Y11);
\draw[very thick,->](eY) to (Y10);
\draw[very thick,->](eY) to (Y00);
\draw[very thick, ->](eY) to (Y01);
\draw[very thick, ->](eM) to (M0);
\draw[very thick, ->](eM) to (M1);

\end{tikzpicture}
\end{center}
\caption{Expansion of Figure \ref{simpledag} to include counterfactuals and possible unmeasured confounders, adapted from Figure 11 in Robins \& Richardson (2011).  $U$ is assumed to be related to both $M$ and $Y$, but only across worlds. Thick arrows denote deterministic relationships.}
\label{RobinsRichardsonFig11}
\end{figure}
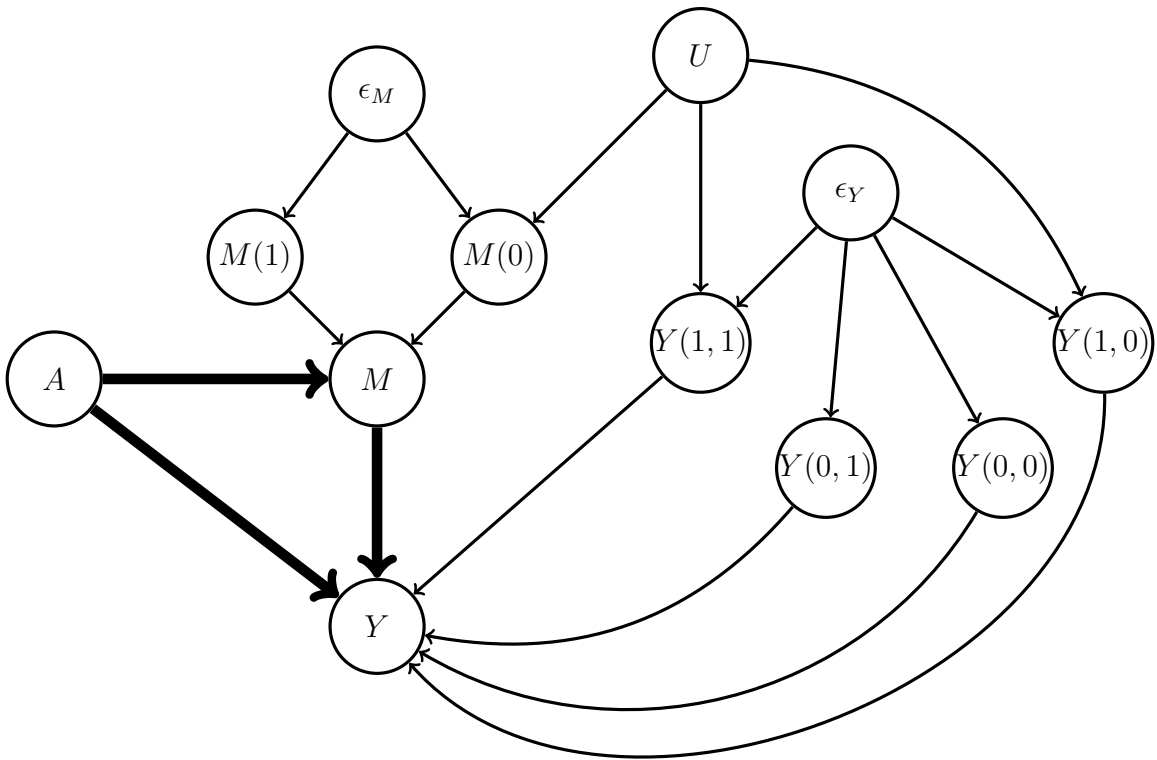

\begin{figure}[ht]
\begin{center}
\begin{tikzpicture}
\tikzstyle{format} = [draw, very thick, circle, minimum size=1.25cm, inner sep=2pt]

\node[format](A){$A$};
\node[format, right=3cm of A, draw=gray, dashed](S){$S$};
\node[format, right=3cm of S](M){$M$};
\node[format, below=3cm of M](Y){$Y$};

\node[format, above left=2.5cm of S](S0){$S(0)$};
\node[format, above=3.5cm of S](S1){$S(1)$};
\node[format, above =1cm of M](M1){$M(1)$};
\node[format, above right=2cm of M](M0){$M(0)$};
\node[format,  right=2cm of M0](U){$U$};
\node[format, below =2cm of Y](Y01){$Y(0,1)$};
\node[format, below right = 2cm of Y](Y00){$Y(0,0)$};
\node[format, above right =2cm of Y](Y11){$Y(1,1)$};
\node[format, right =2cm of Y](Y10){$Y(1,0)$};
\node[format, below right=1.5cm of Y10](eY){$\epsilon_Y$};
\node[format, above right =1cm of M1](eM){$\epsilon_M$};
\node[format, above =1.5cm of S0](eS){$\epsilon_S$};

\draw[very thick, line width=4pt, ->](A) to (S);
\draw[very thick, line width=4pt, dashed, ->](S) to (M);
\draw[very thick, line width=4pt, ->](S) to (Y);
\draw[very thick, bend left, line width=4pt, ->](A) to (M);
\draw[very thick, line width=4pt, ->](M) to (Y);
\draw[very thick, line width=4pt, ->](A) to (Y);
\draw[very thick,->](M0) to (M);
\draw[very thick,->](M1) to (M);
\draw[very thick,->](S0) to (S);
\draw[very thick,->](S1) to (S);
\draw[very thick,->](eS) to (S0);
\draw[very thick,->](eS) to (S1);
\draw[very thick,->](S1) to (S);
\draw[very thick,->](S0) to (S);
\draw[very thick,->](eM) to (M0);
\draw[very thick,->](eM) to (M1);
\draw[very thick,->](Y11) to (Y);
\draw[very thick,->](Y10) to (Y);
\draw[very thick,->](Y01) to (Y);
\draw[very thick,->](Y00) to (Y);
\draw[very thick, bend right, ->](eY) to (Y11);
\draw[very thick,->](eY) to (Y10);
\draw[very thick,->](eY) to (Y00);
\draw[very thick, bend left, ->](eY) to (Y01);

\draw[very thick,->](U) to (M0);
\draw[very thick, bend right, ->](U) to (S1);

\end{tikzpicture}
\end{center}
\caption{Expanded version of Figure \ref{RobinsRichardsonFig11} to fit our knee surgery example, where the question of interest centers on the effect of knee replacement surgery ($A$) on quality of life ($Y$) possibly through walking speed ($M$), where the latent genetic marker ($U$) affects $M$ in the absence of surgery, and affects $Y$ in the presence of surgery through its effect on (unobserved) surgical scarring ($S$). The potential outcomes shown are $S(a)$, $M(a)=M(a,S(a))$, and $Y(a,m)=Y(a,S(a),m)$, $a,m\in \{0,1\}$. The $S\rightarrow M$ edge is dotted to indicate that its presence or absence is important for the the identifying assumptions (see main text for details).}
\label{KneeIntermediateC}
\end{figure}
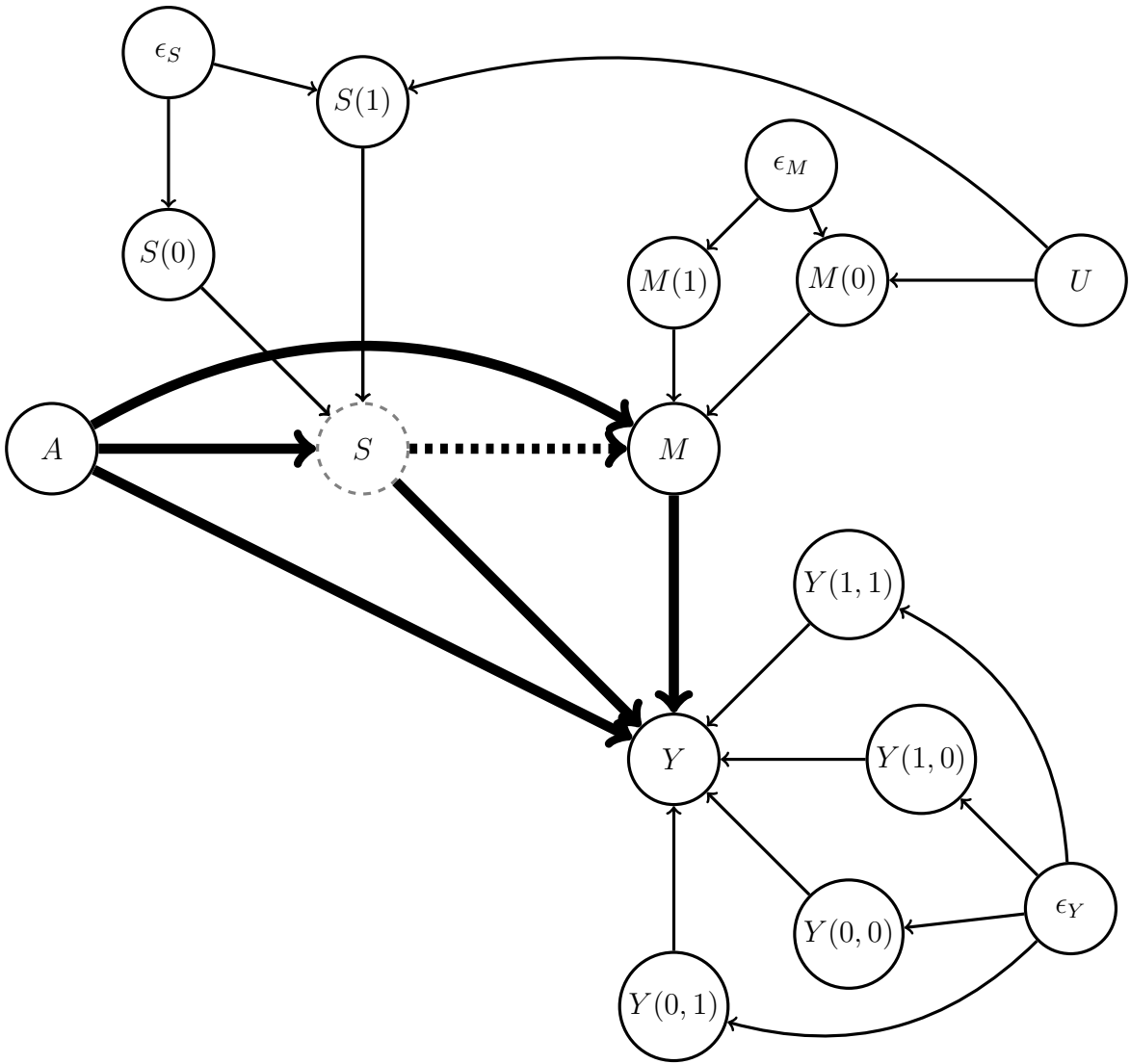

\begin{figure}[t!]
    \centering

        \includegraphics[width=\textwidth,height=\textheight,keepaspectratio]{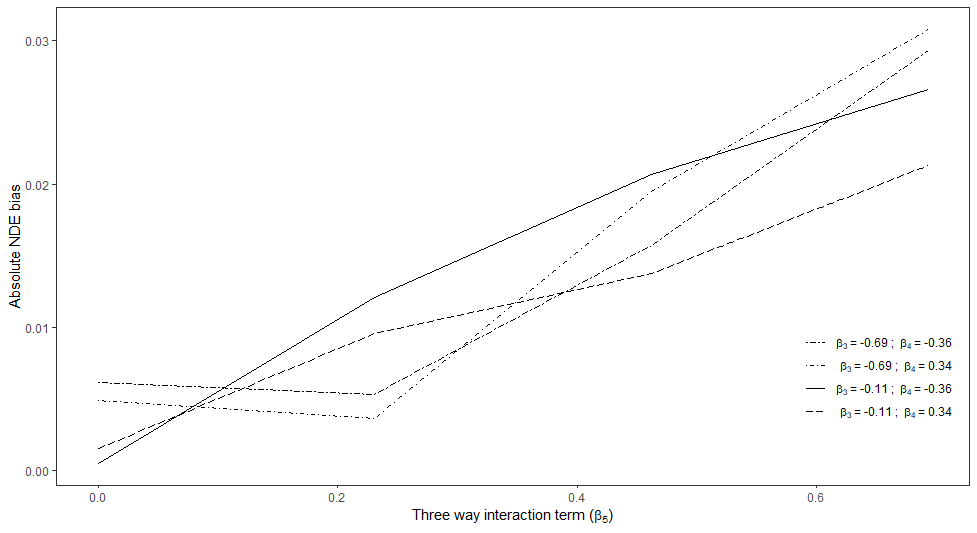}

    \caption{Bias in the g-formula estimand $\widehat{NDE}$ for binary $Y$ as a function of a three-way interaction parameter $\beta_5$ between the exposure ($A$), the mediator ($M$), and an unobserved cross-world confounder ($U$). The bias was calculated using the $\alpha_0$, $\alpha_1$, $\alpha_2$, $\beta_0$, $\beta_1$, and $\beta_2$ values that produced the largest absolute NDE bias, while parameter values for $\beta_3$ and exposure-mediator interaction $\beta_4$ in the outcome model were allowed to take on their minimum and maximum values.}
\label{BiasPlot}
\end{figure}

\clearpage

\newpage

\section*{\center{Supplementary materials}}

\section*{Mathematical justification}

\subsection*{Linear model for continuous $Y$}

As there are no unobserved confounders in Figure 3 in a single world, $\mathbb{E}\{Y(a,M(a)\}$  and  $\mathbb{E}\{M(a)\} $  are identified as $\mathbb{E}\{Y | A=a\}$ and $\mathbb{E}\{M | A=a\}$, respectively, $a=\{0,1\}$. However, because there is unobserved confounding between $M(0)$ and $Y(1,m)$, $\mathbb{E}\{Y(1,M(0))\}$ is not generally equal to $\sum_m \mathbb{E}\{Y \mid A=1, M=m\}p(M=m \mid A=0)$. Specifically, under the structural model of the main text, we have
\begin{eqnarray}
\mathbb{E}\{Y(1,M(0))\} &=& \beta_0 + \beta_1 + (\beta_2 + \beta_4) \mathbb{E}\{M(0)\} + \beta_3 \mathbb{E}\{U\} +  \beta_5 \mathbb{E}\{UM(0)\} + \mathbb{E}\{\epsilon_Y\} \nonumber \\
&=& \beta_0 + \beta_1 + (\beta_2 + \beta_4)\mathbb{E}\{M(0)\} + 2 \beta_3  + \beta_5 \mathbb{E}\{UM(0)\}, \nonumber 
\end{eqnarray}
where
$$
\mathbb{E}\{M(0)\} = p(M=1 \mid A=0) = \mathbb{E}\left\{ \frac{\exp(\alpha_0 + \alpha_2 U)}{1 + \exp(\alpha_0 + \alpha_2 U)} \right\} = \gamma
$$
for some positive constant $\gamma$, and
$$
\mathbb{E}\{UM(0)\} = \psi
$$
for some positive constant $\psi$. Note that because $M(0) \not\independent U$ we cannot assume that $\mathbb{E}\{UM(0)\} = \mathbb{E}\{U\}\mathbb{E}\{M(0)\}$ , which implies that typically $\psi \neq 2\gamma$.

\bigskip

The g-formula estimand $\mathbb{E}\{Y(1,M(0))\}$ is given by \cite{vanderweele2017}:
\begin{eqnarray}
\widehat{\mathbb{E}}\{Y(1,M(0))\} &=& \sum_m \mathbb{E}\{Y \mid A=1, M=m\}p(M=m \mid A=0), \nonumber
\end{eqnarray}
where 
\begin{eqnarray}
\mathbb{E}\{Y \mid A=1, M=m\} &=& \int_u \mathbb{E}\{Y \mid A=1, M=m, U=u\}p(U=u \mid A=1, M=m) du \nonumber \\
 &=& \int_u \mathbb{E}\{Y \mid A=1, M=m, U=u\}p(U=u) du \nonumber \\
&=& \int_u (\beta_0 + \beta_1 +\beta_2 m + \beta_3 u + \beta_4 m + \beta_5 mu)p(U=u) du \nonumber \\
&=& \beta_0 + \beta_1 + (\beta_2 + \beta_4)m + \int_u (\beta_3 u + \beta_5 mu)p(U=u) du \nonumber \\
&=& \beta_0 + \beta_1 + (\beta_2 + \beta_4)m + 2(\beta_3 + \beta_5 m), \nonumber
\end{eqnarray}
where the second line follows from $U \independent M \mid A=1$  and $U \independent A$ by construction. 

\bigskip

With $p(M=1 \mid A=0) =  \gamma$ it follows that
\begin{eqnarray}
\widehat{\mathbb{E}}\{Y(1,M(0))\} &=& (1-\gamma)(\beta_0 + \beta_1 + 2 \beta_3) + \gamma (\beta_0 + \beta_1 + \beta_2 + \beta_4+ 2(\beta_3 + \beta_5))  \nonumber \\
&=& \beta_0 + \beta_1 + 2 \beta_3 + \gamma(\beta_2 + \beta_4 + 2\beta_5) \nonumber 
\end{eqnarray}

\bigskip

Comparing $\mathbb{E}\{Y(1,M(0))\}$ and $\widehat{\mathbb{E}}\{Y(1,M(0))\}$, we see that they  differ by $\beta_5(\psi -2 \gamma)$, which means that as long as the three-way interaction is non-zero, there is a some bias due to cross-world confounding. Hence, we only consider data generating processes where $\beta_5\not= 0$.

\subsection*{Nonlinear model for binary Y}

Assuming that $Y$ follows a logistic structural equation, we have:
\begin{eqnarray}
\mathbb{E}\{Y(1,M(0)\} &=& \int_u \bigg\{ \text{expit} \left( \beta'_0 + \beta'_1 + \beta'_3 u \right)\left(1 - \text{expit}\left(\alpha_0 + \alpha_2 u \right) \right)  \nonumber \\
&+& \text{expit} \left( \beta'_0 + \beta'_1 + \beta'_2 + (\beta'_3+\beta'_5)u + \beta'_4 \right) \left( \text{expit}(\alpha_0 + \alpha_2 u)\right) \bigg\} p(U=u) du \nonumber \\
&=& \eta \nonumber
\end{eqnarray}
for some positive constant $\eta$.

\bigskip

As before $p(M=1 \mid A=0) = \gamma$, and 
\begin{eqnarray}
\mathbb{E}\{Y \mid A=1, M=m\} &=& \int_u \text{expit}(\beta'_0 + \beta'_1 + (\beta'_2 + \beta'_4 + \beta'_5u)m + \beta'_3 u)p(U=u) du. \nonumber 
\end{eqnarray}
Therefore, the g-formula estimand is given by
\begin{eqnarray}
\widehat{\mathbb{E}}\{Y(1,M(0))\} &=& (1-\gamma) \int_u \text{expit} (\beta'_0 + \beta'_1 + \beta'_3 u) p(U=u) du \nonumber \\
&+& \gamma \int_u \text{expit}(\beta'_0 + \beta'_1 + \beta'_2 + \beta'_3u + \beta'_4 + \beta'_5u)p(U=u) du \nonumber \\ 
&=& \eta' \nonumber
\end{eqnarray}
for some positive constant $\eta'$. Because $\eta$ is not necessarily equal to $\eta'$, there is a difference of $\eta - \eta'$ between $\mathbb{E}\{Y(1,M(0))\}$ and $\hat{\mathbb{E}}\{Y(1,M(0))\}$. In this non-linear case, there is a possible bias even when $\beta_5 = 0$; consequently, we evaluated cases in which $\beta_5 = 0$ and cases in which $\beta_5 \neq 0$ in the binary outcome model.

\section*{Choice of parameter values in the numerical example}

We made the following additional assumptions about the data generating mechanisms regarding $M$ and $Y$:

\begin{enumerate}
\item $p(M=1 \mid A=0, U=0)$ is between 0.30 and 0.80, so $\alpha_0$ is between approximately -0.85 and 1.39. 
\item Among those with the same value of $U$, individuals who undergo surgery have 0.70 to 2.50 times the odds of having a normal walking speed compared to those who do not undergo surgery, i.e., $OR_{M,A \mid U} \in [0.70, 2.50]$. Therefore, plausible $\alpha_1$ values are approximately -0.36 to 0.92. 
\item $OR_{M,U \mid A}$ is between 0.30 and 0.90, so plausible $\alpha_2$ values range from approximately -1.20 to -0.11
\end{enumerate}

For the continuous $Y$ case:

\begin{enumerate}
\item $\mathbb{E}(Y\mid A=0, M=0, U=0)$ is between 40 and 60, i.e., $\beta_0 \in [40, 60]$
\item Surgery can either decrease or increase quality of life by up to 10 points, i.e., $\beta_1 \in [-10,10]$
\item A slow walking speed can either decrease quality of life by a maximum of 20 points or increase quality of life by a maximum of 10 points, i.e., $\beta_2 \in [-20, 10]$
\item $U$ decreases quality of life by 5-15 points, i.e., $\beta_3 \in [-15,-5]$
\item There is a negative interaction between $A$ and $M$ on $Y$, i.e., $\beta_4 \in [-20,-10]$
\item There is a moderate three-way interaction between $A$, $M$, and $U$ on $Y$, but the direction is unknown, i.e., $\beta_5 \in [-15,15]$
\end{enumerate}

For the binary $Y$ case:

\begin{enumerate}
\item $p(Y=1 \mid A=0, M=0, U=0)$ is between 0.30 and 0.60, i.e., $\beta_0 \in  [-0.85 , 0.41]$
\item $OR_{Y,A\mid M,U}$ is between 0.50 and 3.0, i.e.,  $\beta_1 \in [-0.22 , 0.90]$
\item $OR_{Y,M \mid A,U}$ is between 1.0 and 3.50, i.e., $\beta_2 \in [0, 1.25]$ 
\item $OR_{Y,U \mid A,M}$ is between 0.50 and 0.90, i.e., $\beta_3 \in [-0.70, -0.10]$
\item There is an interaction between $A$ and $M$ on $Y$. The OR for this interaction term varies between 0.70 and 1.40, i.e., $\beta_4 \in [-0.36, 0.34]$
\item There is a possible three-way interaction between $A$, $M$, and $U$ on $Y$. When present, the OR for this interaction term varies between 1.0 and 2.0, i.e., when non-zero, $\beta_5 \in [0, 0.69]$
\end{enumerate}

\section*{Pseudocode}
The true values of the NDEs and NIEs for the above models and settings, as well as the true values of the g-formula estimands $\widehat{NDE}$ and $\widehat{NIE}$ can be obtained by numerical integration or can be approximated by simulating a huge sample from these models, including the different counterfactual worlds. 
We opt for the latter case and proceed as follows:
\begin{enumerate}
\item Generate one million observations, drawing each $A_i$ and $U_i$ from the following:
\begin{eqnarray*}
A_i \sim \text{Bernoulli}(0.5)\\
U_i \sim \text{Normal}(2,1);
\end{eqnarray*}
\item Generate the corresponding counterfactuals $M_i(0)$, $M_i(1)$, $Y_i(1,M(1))$, $Y_i(1,M(0))$, and $Y_i(0,M(0))$ according to the structural models given in the main text.
\item Obtain the (approximately) true values of the NDE and NIE, 
\begin{eqnarray*}
{\text{NDE}} = \frac{1}{n}\sum (Y_i(1,M(0)) - Y_i(0,M(0)))\\
{\text{NIE}} = \frac{1}{n}\sum (Y_i(1,M(1)) - Y_i(1,M(0)));
\end{eqnarray*}
\item Using consistency, obtain the factual variables $M_i$ and $Y_i$;
\item Calculate the (approximately) true values of the estimands $\widehat{\text{NDE}}$ and $\widehat{\text{NIE}}$ via the mediational g-formula, as one would do with observed data and no knowledge of $U$ 
$$
\widehat{\text{NDE}} = \left(\sum_m \hat{\mathbb{E}}\{Y \mid A=1, M=m\}\hat{p}(M=m \mid A=0)\right) - \widehat{\mathbb{E}}\{Y \mid A=0\}$$
$$\widehat{\text{NIE}} = \widehat{\mathbb{E}}\{Y \mid A=1\} - \left(\sum_m \hat{\mathbb{E}}\{Y \mid A=1, M=m\}\widehat{p}(M=m \mid A=0)\right), $$
where $\hat{\mathbb{E}}\{\cdot\}$ and $\hat{p}(\cdot)$ denote the sample averages and relative frequencies in the particular sub-groups, respectively.
\item The bias for the NDE and NIE is then given by:
$$\Delta_{NDE} ={\text{NDE}}- \widehat{\text{NDE}}\quad \quad \quad \quad \quad \Delta_{NIE}={\text{NIE}}- \widehat{\text{NIE}}.$$
\end{enumerate}

\section*{Background on alternative estimands}

In this section, we expand on the discussion in the main text on alternative estimands by providing additional information and describing how these estimands would (or would not) solve the issue of cross-world independence assumption violations.

\bigskip

We define the {\bf standardized} \cite{geneletti2007identifying} 
(also known as {\bf interventional} \cite{vanderweele2014effect,vansteelandt2017interventional}, or {\bf organic} \cite{lok2016defining}) direct and indirect effects as follows (strictly speaking there are subtle differences between the precise definitions, but these are not essential to the basic ideas of these concepts so we omit the details). Let $M\sim p_a$ denote some intervention which renders $M$ independent of any graph parents (i.e., direct causes of $M$) and generates its value from the distribution of $M(a)$; this could also be a conditional distribution, $p(M(a)|C)$, given some baseline covariates $C$. The standardized (in)direct effects are then defined as
$$
DE_{st}={\mathbb{E}}\{Y(a,M\sim p_{a'})\}-{\mathbb{E}}\{Y(a',M\sim p_{a'})\}
$$ 
and 
$$
IE_{st}={\mathbb{E}}\{Y(a,M\sim p_a)\}-{\mathbb{E}}\{Y(a,M\sim p_{a'})\},
$$
where the expectation is also over that random value of $M$. Note that $DE_{st}$ is a weighted average of the controlled direct effect ($CDE$).

\bigskip

While it is obvious that $DE_{st}+IE_{st}={\mathbb{E}}\{Y(a,M\sim p_a)\}-{\mathbb{E}}\{Y(a',M\sim p_{a'})\}$, it is not necessarily the case that ${\mathbb{E}}\{Y(a,M\sim p_a)\}={\mathbb{E}}\{Y(a)\}$ and hence they do not necessarily add up to the total effect $TE={\mathbb{E}}\{Y(a)\}-{\mathbb{E}}\{Y(a')\}$ \cite{didelez2006direct}.  
Note that Lok \cite{lok2016defining} defines $DE_{org}={\mathbb{E}}\{Y(a,M\sim p_{a'})\}-{\mathbb{E}}\{Y(a')\} $ and 
$IE_{org}={\mathbb{E}}\{Y(a)\}-{\mathbb{E}}\{Y(a,M\sim p_{a'})\}$. The equality  ${\mathbb{E}}\{Y(a,M\sim p_a)\}={\mathbb{E}}\{Y(a)\}$ does hold when property (6) in the main text is satisfied, i.e., $Y(a,m) \independent M(a) \mid (A=a, C)$,  provided that the same baseline covariates $C$ are chosen for conditioning in $p_a=p(M(a)|C)$. In other words, if we want $DE_{st}+IE_{st}=TE$ then we are not entirely free to choose the interventional distribution $p_a$. Hence, the {\it definition, not just the identification}, of the interventional effects may depend on whether and what $M$-$Y$ confounding there is. 
However, it follows from Lok \cite{lok2016defining} that the organic direct and indirect effect do not depend on the choice of $C$ as long as $C$ captures all $M$-$Y$ confounding.
As $C$ cannot include post-exposure confounders of $M$ and $Y$ \cite{didelez2006direct}, the issue of $DE_{st}+IE_{st}\not=TE$ remains when there is intermediate confounding, which for some authors was the main reason to advocate the interventional (in)direct effects. In the particular situation of Figure 3 of the main text, there is no single-world $M$-$Y$ confounding (including no intermediate confounding); however, there is very specific cross-world confounding, i.e., we have that $Y(a,m) \independent M(a) \mid A=a$ with $C=\emptyset$. Thus, it does hold that ${\mathbb{E}}\{Y(a,M\sim p_a)\}={\mathbb{E}}\{Y(a)\}$. 

\bigskip

Regardless of their relation to the total effect, the $DE_{st}$ and $IE_{st}$, as defined above, are identified under assumptions (4)-(6) of the main text, possibly after conditioning on covariates which may or may not equal $C$. Identification is then given by the g-formula, specifically by (7) and (8), again possibly with conditioning on those covariates. Hence, no cross-world independence assumption is used. 

\bigskip

{\it Bibliographic notes:} The ``standardized" notions were introduced by Geneletti  \cite{geneletti2007identifying} and revisited with a focus on the above issue by Didelez et al.\ \cite{didelez2006direct}, both using a decision theoretic framework instead of potential outcomes. They were later suggested as alternative estimands, called interventional analogues, in case of intermediate confounding by VanderWeele et al. \cite{vanderweele2014effect}, where ${\mathbb{E}}\{Y(a,M\sim p_a)\}-{\mathbb{E}}\{Y(a',M\sim p_{a'})\}$ is called the ``total causal effect analogue;'' the authors leave it only implicit that the latter does not equal ${\mathbb{E}}\{Y(a)\}-{\mathbb{E}}\{Y(a')\}$ and hence does not equal the total causal effect. The issue is  briefly addressed by Vansteelandt and Daniel \cite{vansteelandt2017interventional}. Lok \cite{lok2016defining} defines the concepts of organic (in)direct effects with the motivation that in many applications the mediator cannot be set to the particular value $M(a')$; however, the issue of intermediate confounding and its solution are not addressed.\\

A slightly different alternative is the notion of {\bf separable} (in)direct effects. As elaborated by Robins and Richardson \cite{robinsrichardson2011} and Robins, Richardson, \& Shpitser (2020) \cite{robins2020interventionist}, separable (in)direct effects provide a more explicit way to motivate why one may be interested in mediation effects, like natural (in)direct effects, by extending the ``story'' to include hypothetical interventions that allow splitting the effect of the exposure along direct and indirect pathways. They point out that this moves the focus from a cross-world concept to a single world concept. It also moves the focus from intervening on $M$ to a new type of intervention on $A$. Thus, separable effects are always single-world concepts, and require only single world assumptions for identification; however, this single-world is a hypothetical one where the exposure of interest can be decomposed as outlined below. \\

The notion of separable (in)direct effects presupposes that the exposure consists of components which, while observationally identical, could in principle be set to different values in a single world, say $A=(A^Y,A^M)$. 
The targets of inference are then
$$
DE_{sep}={\mathbb{E}}\{Y(A^Y=a,A^M=a')\}-{\mathbb{E}}\{Y(A^Y=a',A^M=a')\}
$$ 
and 
$$
IE_{sep}={\mathbb{E}}\{Y(A^Y=a,A^M=a))-{\mathbb{E}}(Y(A^Y=a,A^M=a')\}
$$
Note that when the hypothetical intervention is such that we believe $Y(A^Y=a,A^M=a)=Y(A=a)$, then $DE_{sep}+IE_{sep}=TE$. 
Note also that for the components ($A^Y,A^M$) to actually reflect mediation through $M$ and nothing else we assume graphically that $A^M$ intersects all directed paths from $A$ to $M$, while $A^Y$ intersects all directed paths from $A$ to $Y$ that are not through $M$\cite{robinsrichardson2011, didelez2019defining}. 
(additional work \cite{stensrudJASA, stensrud2020conditional} describes analogous conditions in the context of competing events). \\

Conditions for identifiability have been given in the literature for special cases: (1) for the three node case $(A,M,Y)$  \cite{robinsrichardson2011}, (2) in the case where $Y$ is a survival time and $M$ a longitudinal mediator \cite{didelez2019defining}, and (3) for the case where both $Y$ and $M$ are time-to-events, with $M$ a competing event for the event $Y$ of interest \cite{stensrudJASA}. The identifying functional is essentially again given by the mediational g-formula, adapted to the dynamic setting where necessary. Assuming a randomized treatment $A$, the identifying conditions for separable effects are still concerned with $M$-$Y$ confounding. This is because the distribution of $Y(A^Y=a,A^M=a')$ needs to be assessed from observational data where  $A=A^M=A^Y$, which violates positivity because we never observe $A^Y=a$ and $A^M=a'$. Informally, there needs to be sufficient information in observed covariates so as to ``separate'' observationally the distribution of $Y$ from $A^M$ given $M$ and these covariates.
Consider Figure 3 in the main text as a purely technical example (i.e., ignore any subject matter context) and assume that  interventions exist such that $A\rightarrow M$ can be replaced by  $A\rightarrow A^M \rightarrow M$, and $A\rightarrow Y$ by  $A\rightarrow A^Y \rightarrow Y$. Further assume that no other edges are into or out of $A^M, A^Y$. The variable $U$ (if unobservable) then renders the separable effects non-identifiable, as $Y\not\independent A^M\mid (M,A^Y=1)$. Note, however, that $U$ is not strictly a confounder for the effects of interest, as it does not affect the intervention nodes  $A^M,A^Y$ (due to randomized $A$). Rather, it prevents identification of the separable effects from observational data where $A=A^M=A^Y$. \\

In the case of intermediate confounding such as Figure 2 in the main text, we find that the separable effects are not well-defined in the sense of reflecting mediation through $M$ and nothing else. In such a case, we need to decide whether $L$ is directly affected by component $A^M$ or $A^Y$, by both, or something else like a third component (see \cite{robinsrichardson2011,stensrud2020conditional}). Hence, the case of intermediate confounding requires a redefinition of the relevant separable effects of interest, and these will depend on the subject matter in any given practical application. \\

In the context of our knee surgery example, to justify separability we would need to imagine a component of knee surgery which acts on walking speed like a knee replacement does but without affecting quality of life through any other pathways, and a second component which acts on quality of life (including knee scarring) like surgery but without affecting walking speed. Further, we assume that, at least hypothetically, these components can separately be set to different values. While there are clearly different separable components to knee surgery (e.g., quality of the artificial knee versus the quality of the surgeon),
it is still difficult to think of a realistic practical intervention that reflects exactly the notion of separable effects with $M$ being walking speed, as this will likely be affected by all of these components.  \\

{\it Bibliographic notes:} The principle of separable effects  was introduced by Robins and Richardson (2011) \cite{robinsrichardson2011}, without actually calling them ``separable;'' they show that NDE and NIE are never identified under a FFRCISTG, but that $DE_{sep}$ and $IE_{sep}$ are identified under a FFRCISTG given an augmented DAG. In Robins \& Richardson (2013) \cite{richardson2013swig} they also show that these concepts are compatible with SWIGs. Didelez \cite{didelez2019defining} pointed out that NDE and NIE are not well-defined when $Y$ is survival and $M$ is longitudinal, while separable effects are. Aalen et al. \cite{aalen2019time} propose estimation based on the additive hazard model and present a practical application. A series of papers by Stensrud et al. \cite{stensrudJASA, stensrud2020generalized,stensrud2020conditional} actually introduce the name ``separable effects'', and  extended these ideas to the situation of time-to-event settings with competing events; they give formal definitions, conditions for identification and estimation procedures, as well as an application. Even more recently, Robins, Richardson, \& Shpitser (2020) \cite{robins2020interventionist} expand upon the original Robins and Richardson (2011) work to develop an ``interventionist"' approach to causal mediation. \\

In summary, we believe that alternative estimands could be suitable depending on what assumptions one is willing (or unwilling) to make. Each approach has its own strengths and weaknesses; there is no universally superior approach in general. Above all, it is crucial that the applied researcher think carefully about his or her research question and desired target of interest to ensure that the most suitable estimand is chosen.

\clearpage
\newpage

\section*{R code: Linear $Y$}

\begin{lstlisting}
library(dplyr)
library(purrr)
library(faraway)
library(foreign)

alpha0_vec = seq(logit(0.3), logit(0.8), length.out=4)
alphaA_vec = seq(log(0.7), log(2.5), length.out=4)
alphaU_vec = seq(log(0.3), log(0.9), length.out=4)

beta0_vec = seq(40, 60, length.out=4)
betaA_vec = seq(-10, 10, length.out=4)
betaM_vec = seq(-20, 10, length.out=4)
betaU_vec = seq(-15,-5, length.out=4)
betaAM_vec =seq(-20,-10, length.out=4)
betaMU_vec = seq(-15,15, length.out=4)

#Set up dataframe for the grid search


grids <- list(alpha0 = alpha0_vec,
              alphaA = alphaA_vec,
              alphaU = alphaU_vec,
              beta0 = beta0_vec,
              betaA = betaA_vec,
              betaM = betaM_vec,
              betaU = betaU_vec,
              betaAM = betaAM_vec,
              betaMU = betaMU_vec) \%>\% cross_df()

#Specify N
n = 1e6



genM = function(n, trt, unobs, alpha0, alphaA, alphaU){
  eqnM = alpha0 + alphaA*trt + alphaU*(1-trt)*unobs
  probM = ilogit(eqnM)
  M = rbinom(n, size=1, prob=probM)
  return(M)
}

genY_bin = function(n, trt, med, unobs, beta0, betaA, betaM, 
betaU, betaAM, betaMU){
  eqnY = beta0 + betaA*trt + betaM*med +
   betaU*(trt)*unobs + betaAM*trt*med 
   + betaMU*trt*med*unobs
  probY = ilogit(eqnY)
  Y = rbinom(n, size=1, prob=probY)
  return(Y)
}

genY_cont = function(n, trt, med, unobs, beta0, betaA, betaM,
 betaU, betaAM, betaMU){
  eqnY = beta0 + betaA*trt + betaM*med 
  + betaU*(trt)*unobs + betaAM*trt*med 
  + betaMU*trt*med*unobs
  Y = rnorm(n=n, mean=eqnY, sd=sqrt(10))
  return(Y)
}


set.seed(2470)

#Generate exogenous variables
U = rnorm(n, mean=2, sd=1)

num_searches = dim(grids)[1]

pM0 = rep(0, num_searches)
pM1 = rep(0, num_searches)
pY00 = rep(0, num_searches)
pY10 = rep(0, num_searches)
pY11 = rep(0, num_searches)

trNDE = rep(0, num_searches)
trNIE = rep(0, num_searches)
epNDE = rep(0, num_searches)
epNIE = rep(0, num_searches)

nde_diff = rep(0, num_searches)
nie_diff = rep(0, num_searches)

for (i in 1:num_searches){
  params = as.numeric(grids[i,])
  M0 <- genM(n, trt=0, unobs = U,
             alpha0 = params[1],
             alphaA = params[2],
             alphaU = params[3])
  
  M1 <- genM(n, trt=1, unobs = U,
             alpha0 = params[1],
             alphaA = params[2],
             alphaU = params[3])
  
  pM0[i] <- mean(M0)
  pM1[i] <- mean(M1)

  Y_A0M0 = genY_cont(n, trt=0, med=M0, unobs=U,
                    beta0 = params[4],
                    betaA = params[5],
                    betaM = params[6],
                    betaU = params[7],
                    betaAM = params[8],
                    betaMU = params[9])

  Y_A1M0 = genY_cont(n, trt=1, med=M0, unobs=U,
                    beta0 = params[4],
                    betaA = params[5],
                    betaM = params[6],
                    betaU = params[7],
                    betaAM = params[8],
                    betaMU = params[9])

  Y_A1M1 = genY_cont(n, trt=1, med=M1, unobs=U,
                    beta0 = params[4],
                    betaA = params[5],
                    betaM = params[6],
                    betaU = params[7],
                    betaAM = params[8],
                    betaMU = params[9])
  
  pY00[i] <- mean(Y_A0M0)
  pY11[i] <- mean(Y_A1M1)
  pY10[i] <- mean(Y_A1M0)
  
  EY_A1M0 = mean(Y_A1M0)
  EY_A0M0 = mean(Y_A0M0)
  NDE = EY_A1M0 - EY_A0M0
  trNDE[i] <- NDE
  
  EY_A1M1 = mean(Y_A1M1)
  NIE = EY_A1M1 - EY_A1M0
  trNIE[i] <- NIE
  
  #Generate A
  A <- rbinom(n, size=1, prob=0.5)
  
  #Generate M using same model as simulation 1,
  # but let trt input be A
  M = genM(n, trt=A, unobs=U, 
           alpha0 = params[1],
           alphaA = params[2],
           alphaU = params[3])
  
  #Generate Y using same model as simulation 1, but let trt input
  # be A and med input be M
  Y = genY_cont(n, trt=A, med=M, unobs=U,
               beta0 = params[4],
               betaA = params[5],
               betaM = params[6],
               betaU = params[7],
               betaAM = params[8],
               betaMU = params[9])
  
  #Create a dataset based on A, M, and Y
  d = as.data.frame(cbind(A,M,Y))
  
  #g-formula for E(Y(1,M(0))

  hatEY_A1M0 = mean(d$Y[d$A==1 & d$M==0])
  hatEY_A1M1 = mean(d$Y[d$A==1 & d$M==1])
  hatPM1_A0 = mean(d$M[d$A==0])
  
  gform_EY_A1M0 = hatEY_A1M0*(1-hatPM1_A0) +
   hatEY_A1M1*hatPM1_A0
  
  #Calculate E(Y(1,M(1))) & E(Y(0,M(0)))
  obs_EY_A1M1 = mean(d$Y[d$A==1])
  obs_EY_A0M0 = mean(d$Y[d$A==0])
  
  obs_NDE = gform_EY_A1M0-obs_EY_A0M0
  epNDE[i] <- obs_NDE
  
  obs_NIE = obs_EY_A1M1-gform_EY_A1M0
  epNIE[i] <- obs_NIE
  

  nde_diff[i] <- NDE - obs_NDE
  nie_diff[i] <- NIE - obs_NIE
  
}

simResult <- mutate(grids,
                pM0=pM0,
                pM1 = pM1,
                pY11 = pY11,
                pY10 = pY10,
                pY00 = pY00,
                nde_diff = nde_diff,
                nie_diff = nie_diff,
                trNDE = trNDE,
                trNIE = trNIE,
                epNDE = epNDE,
                epNIE = epNIE)

\end{lstlisting}

\clearpage
\newpage

\section*{R code: Nonlinear Y}

\begin{lstlisting}

library(foreign)
library(dplyr)
library(purrr)

#Specify N
n = 1e6



genM = function(n, trt, unobs, alpha0, alphaA, alphaU){
  eqnM = alpha0 + alphaA*trt + alphaU*(1-trt)*unobs
  probM = ilogit(eqnM)
  M = rbinom(n, size=1, prob=probM)
  return(M)
}

genY_bin = function(n, trt, med, unobs, beta0, betaA,
 betaM, betaU, betaAM, betaMU){
  eqnY = beta0 + betaA*trt + betaM*med + betaU*(trt)*unobs + 
  betaAM*trt*med + betaMU*trt*med*unobs
  probY = ilogit(eqnY)
  Y = rbinom(n, size=1, prob=probY)
  return(Y)
}



alpha0_vec = seq(logit(0.3), logit(0.8), length.out=4)
alphaA_vec = seq(log(0.7), log(2.5), length.out=4)
alphaU_vec = seq(log(0.3), log(0.9), length.out=4)

beta0_vec = seq(logit(0.3), logit(0.6), length.out=4)
betaA_vec = seq(log(0.5), log(3.0), length.out=4)
betaM_vec = seq(log(1.0), log(3.5), length.out=4)
betaU_vec = seq(log(0.5), log(0.9), length.out=4)
betaAM_vec = c(seq(log(0.7), log(1.4), length.out=4),0)
betaMU_vec = seq(log(1.0), log(2.0), length.out=4)


grids <- list(alpha0 = alpha0_vec,
              alphaA = alphaA_vec,
              alphaU = alphaU_vec,
              beta0 = beta0_vec,
              betaA = betaA_vec,
              betaM = betaM_vec,
              betaU = betaU_vec,
              betaAM = betaAM_vec,
              betaMU = betaMU_vec) \%>\% cross_df()
  



set.seed(2470)

#Generate exogenous variables
U = rnorm(n, mean=2, sd=1)


num_searches = dim(grids)[1]

pM0 = rep(0, num_searches)
pM1 = rep(0, num_searches)
pY00 = rep(0, num_searches)
pY10 = rep(0, num_searches)
pY11 = rep(0, num_searches)

trNDE = rep(0, num_searches)
trNIE = rep(0, num_searches)
epNDE = rep(0, num_searches)
epNIE = rep(0, num_searches)

nde_diff = rep(0, num_searches)
nie_diff = rep(0, num_searches)

for (i in 1:20){
  params = as.numeric(grids[i,])
  M0 <- genM(n, trt=0, unobs = U,
             alpha0 = params[1],
             alphaA = params[2],
             alphaU = params[3])
  
  M1 <- genM(n, trt=1, unobs = U,
             alpha0 = params[1],
             alphaA = params[2],
             alphaU = params[3])
  
  pM0[i] <- mean(M0)
  pM1[i] <- mean(M1)

  Y_A0M0 = genY_bin(n, trt=0, med=M0, unobs=U,
                    beta0 = params[4],
                    betaA = params[5],
                    betaM = params[6],
                    betaU = params[7],
                    betaAM = params[8],
                    betaMU = params[9])

  Y_A1M0 = genY_bin(n, trt=1, med=M0, unobs=U,
                    beta0 = params[4],
                    betaA = params[5],
                    betaM = params[6],
                    betaU = params[7],
                    betaAM = params[8],
                    betaMU = params[9])

  Y_A1M1 = genY_bin(n, trt=1, med=M1, unobs=U,
                    beta0 = params[4],
                    betaA = params[5],
                    betaM = params[6],
                    betaU = params[7],
                    betaAM = params[8],
                    betaMU = params[9])
  
  pY00[i] <- mean(Y_A0M0)
  pY11[i] <- mean(Y_A1M1)
  pY10[i] <- mean(Y_A1M0)
  
  EY_A1M0 = mean(Y_A1M0)
  EY_A0M0 = mean(Y_A0M0)
  NDE = EY_A1M0 - EY_A0M0
  trNDE[i] <- NDE
  
  EY_A1M1 = mean(Y_A1M1)
  NIE = EY_A1M1 - EY_A1M0
  trNIE[i] <- NIE
  
  #Generate A
  A <- rbinom(n, size=1, prob=0.5)
  
  #Generate M using same model as simulation 1,
  # but let trt input be A
  M = genM(n, trt=A, unobs=U, 
           alpha0 = params[1],
           alphaA = params[2],
           alphaU = params[3])
  
  #Generate Y using same model as simulation 1, 
  #but let trt input be A and med input be M
  Y = genY_bin(n, trt=A, med=M, unobs=U,
               beta0 = params[4],
               betaA = params[5],
               betaM = params[6],
               betaU = params[7],
               betaAM = params[8],
               betaMU = params[9])
  
  #Create a dataset based on A, M, and Y
  d = as.data.frame(cbind(A,M,Y))
  
  #g-formula for E(Y(1,M(0))

  
  hatEY_A1M0 = mean(d$Y[d$A==1 & d$M==0])
  hatEY_A1M1 = mean(d$Y[d$A==1 & d$M==1])
  hatPM1_A0 = mean(d$M[d$A==0])
  
  gform_EY_A1M0 = hatEY_A1M0*(1-hatPM1_A0) + hatEY_A1M1*hatPM1_A0
  
  #Calculate E(Y(1,M(1))) & E(Y(0,M(0)))
  obs_EY_A1M1 = mean(d$Y[d$A==1])
  obs_EY_A0M0 = mean(d$Y[d$A==0])
  
  obs_NDE = gform_EY_A1M0-obs_EY_A0M0
  epNDE[i] <- obs_NDE
  
  obs_NIE = obs_EY_A1M1-gform_EY_A1M0
  epNIE[i] <- obs_NIE
  

  nde_diff[i] <- NDE - obs_NDE
  nie_diff[i] <- NIE - obs_NIE
  
}

simResult <- mutate(grids,
                pM0=pM0,
                pM1 = pM1,
                pY11 = pY11,
                pY10 = pY10,
                pY00 = pY00,
                nde_diff = nde_diff,
                nie_diff = nie_diff,
                trNDE = trNDE,
                trNIE = trNIE,
                epNDE = epNDE,
                epNIE = epNIE)


\end{lstlisting}

\clearpage
\newpage

\bibliography{CWbibtex}
\bibliographystyle{epidemiology2}

\end{document}